\documentclass[referee]{raa}            

\usepackage{graphicx,times,natbib}             

\begin{document}

   \title{Double-lined M-dwarf Eclipsing Binaries from Catalina Sky Survey and LAMOST
}

   \volnopage{Vol.0 (200x) No.0, 000--000}      
   \setcounter{page}{1}          

   \author{Chien-Hsiu Lee
      \inst{1}
   \and Chien-Cheng Lin
      \inst{2}
   }

   \institute{Subaru Telescope, National Astronomical Observatory of Japan, 650 North Aohoku Place, Hilo, HI 96720, USA; {\it leech@naoj.org}
             \and
             Shanghai Astronomical Observatory, Chinese Academy of Sciences, 80 Nandan Road, Shanghai 200030, China; {\it cclin@shao.ac.cn}
   }

   \date{Received~~2016 month day; accepted~~20xx~~month day}

   \abstract{Eclipsing binaries provide a unique opportunity to determine fundamental stellar properties. In the era of wide-field camera and all-sky imaging surveys, thousands of eclipsing binaries have been reported by light curve classification, yet their basic properties remain unexplored due to the extensive efforts needed to follow them up spectroscopically. In this paper we investigate three M2-M3 type double-lined eclipsing binaries discovered by cross-matching eclipsing binaries from Catalina Sky Surveys and spectroscopically classified M dwarves from Large Sky Area Multi-Object fibre Spectroscopic Telescope survey data release one and two. Because these three M dwarf binaries are faint, we further acquire radial velocity measurements using GMOS on-boad the Gemini north telecope with R$\sim$4000, enabling us to determine the mass and radius of individual stellar components. By joint fitting of the light and radial velocity curves of these systems, we derive the mass and radius of the primary and secondary components of these three systems, in the range between 0.28-0.42 M$_\odot$ and 0.29-0.67R$_\odot$, respectively. Future spectroscopic observation with high resolution spectrograph will help us pin down the uncertainties of their stellar parameters, and renders these systems benchmarks to study M dwarves, providing inputs to improving stellar models at low mass regime, or establish an empirical mass-radius relation for M dwarf stars.
   }

   \authorrunning{C.-H. Lee \& C.-C. Lin }            
   \titlerunning{M-dwarf EB from CSS$\times$LAMOST}  

   \maketitle

%
%

\section{Introduction}
M dwarves are the cool, low mass end of the main sequence; they are 
the most abundant stars in the Milky Way and ubiquitous in
the solar neighbourhood. However, they are very faint in the optical bands, 
hence are difficult to discover and characterize. Thanks to the advances of large
spectroscopic surveys using medium-class telescopes equipped with wide-filed cameras and multi-object
spectrographs, such as Sloan Digital Sky Survey \citep{2000AJ....120.1579Y} and Large Sky Area Multi-Object fibre Spectroscopic Telescope \citep[LAMOST][]{2012RAA....12.1197C},
hundreds of thousands of M dwarves have been spectroscopically confirmed to-date
\citep{2011AJ....141...97W,2015RAA....15.1182G}.

Recent progresses of M dwarf spectroscopic studies also cast 
challenges to theoretical modeling. For examples, it has been found that
there are descrepencies between  theoretical modeling and observations of M dwarves.
The models often under-predict the size and  over-predict the temperature of M dwarves
\citep[see e.g.][]{2015MNRAS.451.2263Z}, especially in the very low mass end
(M$<$0.3M$_\odot$) where the stars become completely convective and therefore difficult to be modelled. 
Further more, it has been reported that M dwarves exhibit strong magnetic activities, 
which might be linked to their inflated size. Deriving fundamental parameters of 
M dwarves have drawn more and more interests, especially because an increasing number of
exoplanets discovered by {\it Kepler} mission are hosted by M-type stars \citep{2013ApJ...767...95D}. 
For transiting exoplanets, precise and accurate measurements of the host star 
parameters are crucial to derive exoplanets' radius, which is essential to infer the
density of the exoplanets and determine if they are gaseous, Jupiter-like planets or rocky planets
like Earth.
In this regard, improving theoretical modeling of M dwarf stars, or establishing an
emperical mass-radius relation for M dwarves is highly demanded. 
       
Eclipsing binaries provide us unique opportunities to directly  measure the mass,
radius, and effective temperature of individual stars down to a few percents level. Their
light curves can provide information on the inclination angle, orbital period, eccentricity, mass
ratio, and radius in terms of the semi-major axis. On the other hand, spectroscopic observations will
enable us to derive the mass and effective temperature of individual stars, as well as their
orbital distance. Note the normal effective temperature and surface gravity degeneracy of extracting
information from a single star spectrum is not an issue for eclipsing binaries. This is because we can
directly determine the mass and radius and derive the surface gravity independently, hence break the degeneracy and unambiguously determine the temperature of the binaries.

This work reports the discovery of three double-lined M dwarf
eclipsing binary systems by joining the forces of Catalina Sky Surveys and LAMOST survey,
as well as dedicated spectroscopic follow-ups using Gemini telescope.
This paper is organized as follows: in \textsection \ref{sec.data} we describe the photometric and
spectroscopic observation in hand. We present our analysis in 
\textsection \ref{sec.ana}, followed by a discussion \textsection \ref{sec.dis} and prospects in \textsection \ref{sec.sum}. 

\section{Data} \label{sec.data}

\subsection{Known M dwarves from LAMOST}
Because of their faintness in optical wavelengths, previous studies of M dwrves were carried out using infrared imaging \citep[see e.g.][]{2011AJ....142..138L,2013PASP..125..809T}.
Thanks to the advances in wide-field cameras, multi-object spectrographs, and all sky surveys with medium-size telescopes,
it is possible to obtain spectra to confirm M dwarves in quantity and quality.
In this work we make use of the spectral catalog from LAMOST\footnote{www.lamost.org}, a 4-m class telescope equipped with 4000 fibres patrolling a 5 degree field-of-view, with the capability to
deliver spectra covering a wide range of the optical band (3700-9000\AA) with a resolution R=1800 for target as faint as r=19 magnitudes \citep{2012RAA....12.1197C}.
\cite{2015RAA....15.1182G} have made use of
LAMOST data release one, and present a sample of $\sim$ 93,000
spectroscopically confirmed and classified M dwarves. In the most recent LAMOST data release (DR2), there are $\sim$ 220,000 M dwarves
being catalogued, more than doubling the number of M dwarves in LAMOST DR1.
The LAMOST M dwarf sample were classified by the Hammer spectral typing facility \citep{2007AJ....134.2398C}, aided by visual inspection,
making sure that the spectral type is determined as precise as within one subtype. This is by far the largest M dwarf sample confirmed spectroscopically,
providing us a firm basis to cross-matching with eclipsing binaries charted by other time-domain surveys.
           
\subsection{Time series photometry from CSS} \label{sec.lc}

Since 2004, the Catalina Sky Survey (CSS) constantly patrol the sky between -75$<$Dec$<$70 degrees using three small telescopes. To ensure the coverage of most part of the entire sky, two of the telescopes, i.e. the 0.7 m Catalina Schmidt Telescope (with 8.2 deg$^2$ FOV) and the 1.5 m Mount Lemmon Telescope (with 1 deg$^2$ FOV) are in the northern hemisphere in Arizona, USA, while the 0.5 m Siding Spring Telescope (with 4.2 deg$^2$ FOV) observes the southern hemisphere, based in Australia. 
The main scientific driver of CSS is near Earth objects and potentially hazardous asteroids. Each telescope patrols it own patches of sky, excluding the regions with Galactic latitude less than 15 degrees to avoid crowded stellar fields.
The observations were carried out without any filter to maximize the throughput. The observations were taken in groups of four exposures separated by 10 minutes, with individual integration times as long as 30 seconds. After CCD reduction, the images were fed to SExtractor \citep{1996A&AS..117..393B} for apperture photometry. Thanks to the high cadence and wide patroling area, the survey is also valuable for time-domain science, and enables the Catalina Real-time Transient Survey \citep[CRTS][]{2009ApJ...696..870D}.

The data of the 0.7 m Catalina Schmidt Telescope between 2005 and 2011 were made public
through Catalina Surveys Data Release\footnote{http://nesssi.cacr.caltech.edu/DataRelease/}; the first data release (CSDR1) contains more than 200 million sources between 12 and 20 magnitudes in V-band \citep{2013ApJ...763...32D}.
\cite{2014ApJS..213....9D} searched variables in CSDR1 in the region of -22$<$Dec$<$65 degrees, and
identified $\sim$ 47,000 periodic variables using Lomb-Scargle periodgram. They further visually inspected the
light curves and catalogued 4683 detached eclipsing binaries,
the largest all-sky eclipsing binary catalog up-to-date, providing a
wealth resource of time-series photometry.

\begin{figure}[!h]
  \includegraphics[width=\columnwidth]{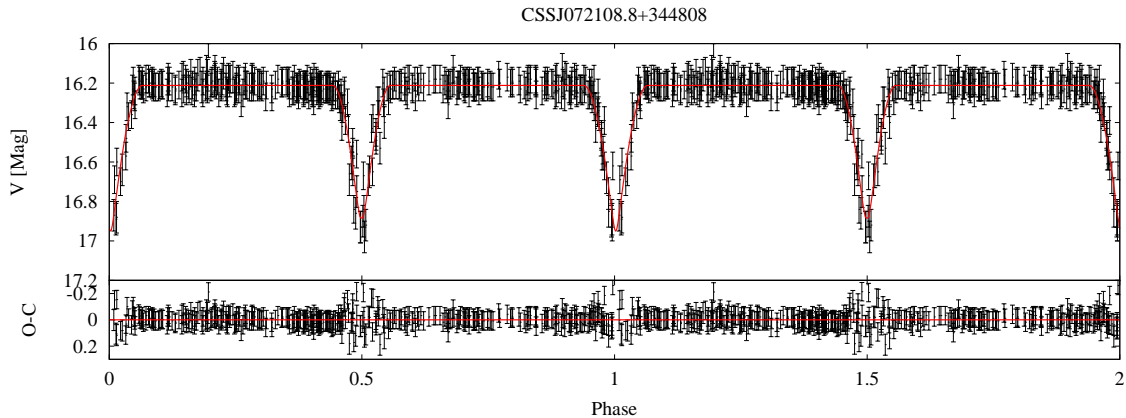}
  \caption{Example Catalina Sky Survey light curve of the M-dwarf binary systems.
    Upper panel: the CSS photometry are marked in black,
    while the best-fit DEBiL model (see \textsection \ref{sec.ana}) is marked in red.
    Lower panel: residuals of the best-fit DEBiL model.}
  \label{fig.lc}
\end{figure}

\section{Analysis}
\label{sec.ana}
\subsection{Light curve analysis using DEBiL}

As a starting point, we use the Detached Eclipsing Binary Light curve fitter\footnote{http://debil.droppages.com} 
\citep[DEBiL,][]{2005ApJ...628..411D} to analyse the light curves of our three M dwarf eclipsing binaries.
DEBiL assumes a simple geometry to model the light curves: it forms an initial guess of the binary configuration
from a set of analytic formula for detached binaries \citep[see e.g.][]{2003ApJ...585.1038S}, and attempts to fit
several parameters:\\
1. The relative radius ($\frac{R_1}{a}$,$\frac{R_2}{a}$), in terms of the semi-major axis $a$.\\
2. The brightness of both stars (mag$_1$,mag$_2$).\\
3. The inclination angle ($i$), eccentricity ($e$), and Argument of periastron ($\omega$) of the binary orbit.\\

DEBiL fits the light curves in an iterative manner by  
minimising $\chi^2$ using the downhill simplex method (Nelder \& Mead, 1965) 
with simulated annealing \citep{1992nrfa.book.....P}. Our best-fit results from DEBiL, as well as the light curve from CSDR1,
are shown in Fig. \ref{fig.lc}. The best-fit parameters from DEBiL are then fed to JKTEBOP for dynamical analysis in Section \ref{sec.dyn}.

\subsection{Radial velocity determination}
\label{sec.spec}
As our eclipsing binaries are faint and their periods are short (P$\sim$0.4 -- 1.3 days), to prevent
smooth-out of the radial velocity (RV) curves, we need to reach sufficient signal-to-noise ratio (S/N), i.e. S/N$>$50 within 0.1 periods,
which can be as short as one hour.
In addition, a sufficient spectral resolution is also required to determine the masses and radii of the binary systems, preferrably less than 1\AA/pixel.
Only 8-m class telescopes are capable to reach such high S/N and spectral resolution
in the given amount of time. We thus conduct spectroscopic follow-up observations using
GMOS \citep{2004PASP..116..425H} on-board Gemini telescope via fast turnaround program (Progam ID GN-2016-FT16).
We use R831 grating with 0.5 arcsecond slit, rendering a resolution R=4396, with central wavelength at 7000\AA
and a wavelength coverage of $\sim$2000\AA.

The observations were carried out on 2016 April 2nd and 13th during the expected radial velocity maxima at
light curve quadrature (phase 0.25 and/or 0.75). In principle we need only one observation at each quadrature
to measure the radial velocity maximum, nevertheless we aim to obtain two exposures at per maxumum, each
with exposure times in 300 (CSSJ092128.3+332558), 400 (CSSJ074118.8+311434), and 600 (CSSJ072108.8+344808) seconds to reach high S/N. The second exposure can serve as a sanity check, ensuring we
obtain consistent radial velocity measurement at the same maximum.
The GMOS data are reduced by using the dedicated IRAF\footnote{http://iraf.noao.edu} GMOS
package\footnote{http://www.gemini.edu/sciops/data-and-results/processing-software} (v1.13) in a standard manner:
each spectrum was bias subtracted, flat fielded, sky subtracted, and wavelength calibrated using CuAr lamp. 

To extract radial velocity information, we make use of the bright H$\alpha$ emission line at 6562.8\AA.
The H$\alpha$ emissions from both stellar component are clearly resolved. We then
fit the H$\alpha$ line profile using a two-component Gaussian function. From the Gaussian fit, we
obtain the radial velocity of each stellar component, as shown in Table \ref{tab.rv}, with estimated error of
15 km/s based on the spectral resolution delivered by the R831 grating and 0.5 arcsecond slit (i.e. 3.4 \AA/pixel).

\begin{table}
\centering
\caption{Radial velocity measurements from GMOS observations.}
\begin{tabular}[t]{lrr}
  \hline
  \hline
  Epoch & RV$_1^\dagger$ & RV$_2^\dagger$ \\
        & [km/s] & [km/s] \\
\hline
\multicolumn{3}{c}{{\it CSSJ074118.8+311434}} \\
57480.230475 & 202.05 &  22.86 \\ 
57480.238484 & 203.42 &  23.77 \\ 
57491.246632 &  56.68 & 195.19 \\ 
\hline
\multicolumn{3}{c}{{\it CSSJ092128.3+332558}} \\
57491.297419 & 138.51 & -92.80 \\ 
57491.301748 & 141.25 & -90.05 \\ 
57491.306088 & 143.99 & -92.80 \\ 
\hline
\multicolumn{3}{c}{{\it CSSJ072108.8+344808}} \\
57480.359583 & 118.85 & -128.45 \\ 
57480.369097 & 117.48 & -131.19 \\ 
\hline
\hline
\multicolumn{3}{l}{$\dagger$ The estimated error is $\pm$ 15 km/s.} 
\end{tabular}
\label{tab.rv}
\end{table}

\subsection{Dynamical analysis} \label{sec.dyn}

\begin{figure*}[!t]
  \centering
  \includegraphics[scale=0.35]{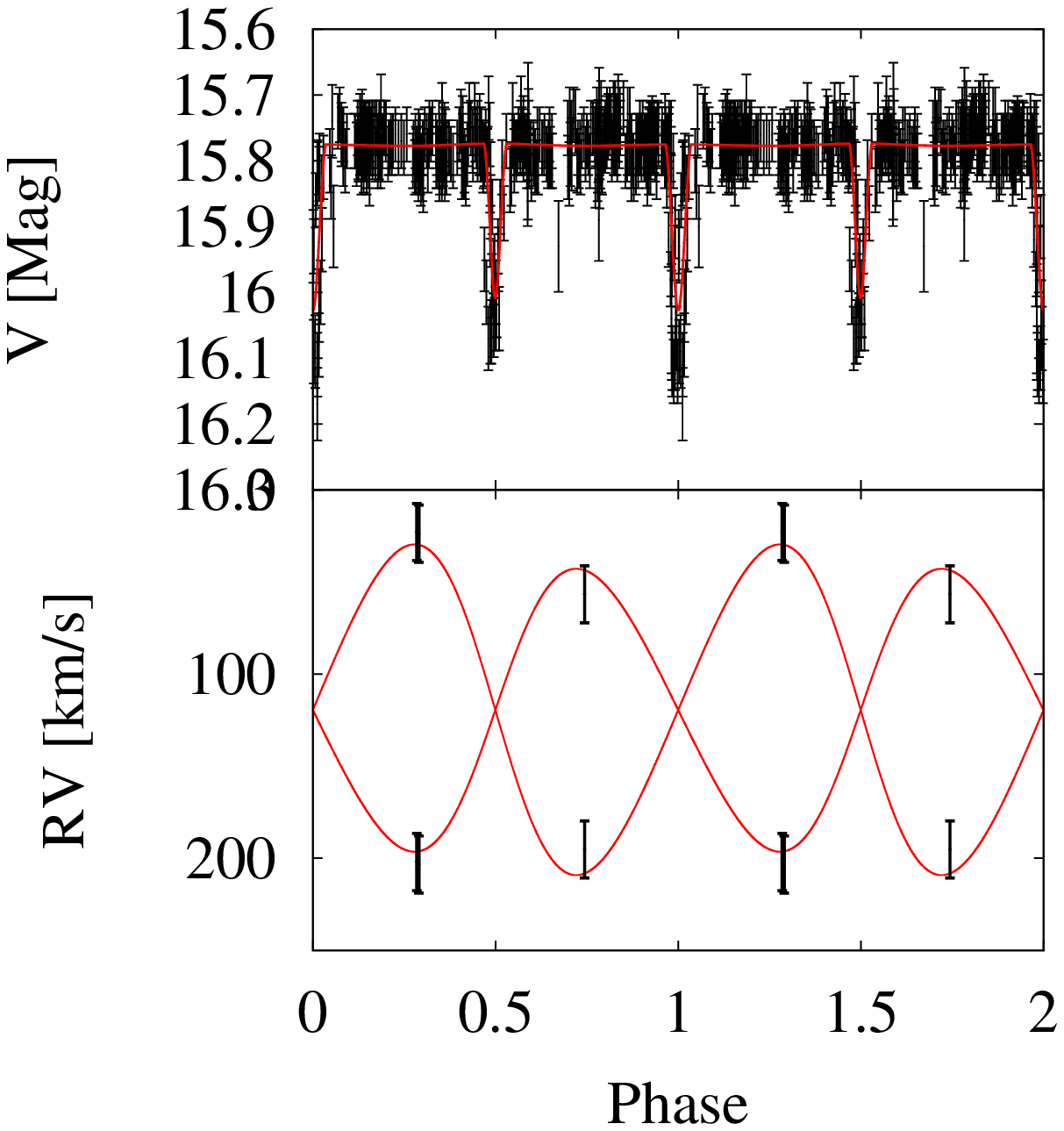}
  \includegraphics[scale=0.35]{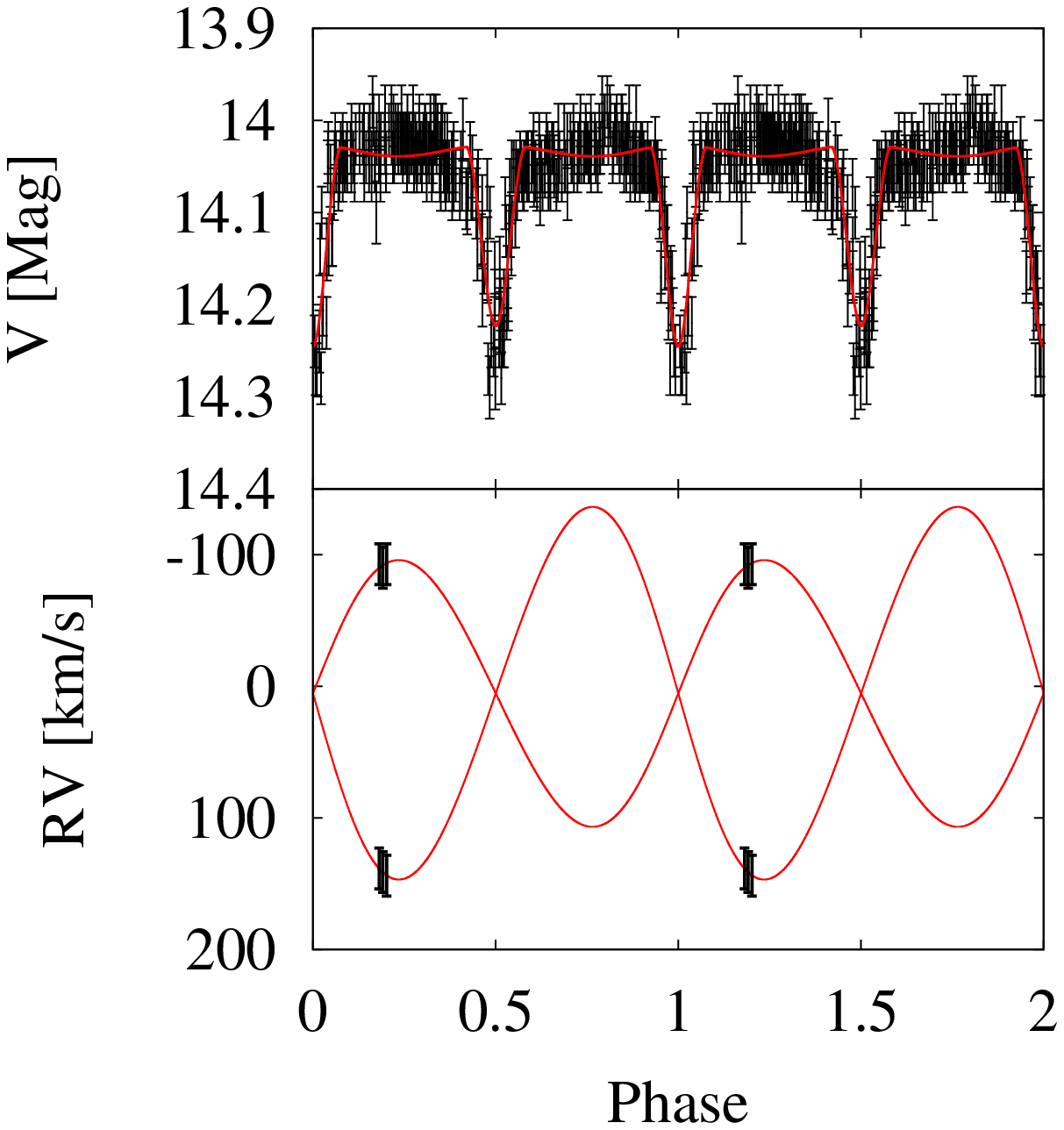}
  \includegraphics[scale=0.35]{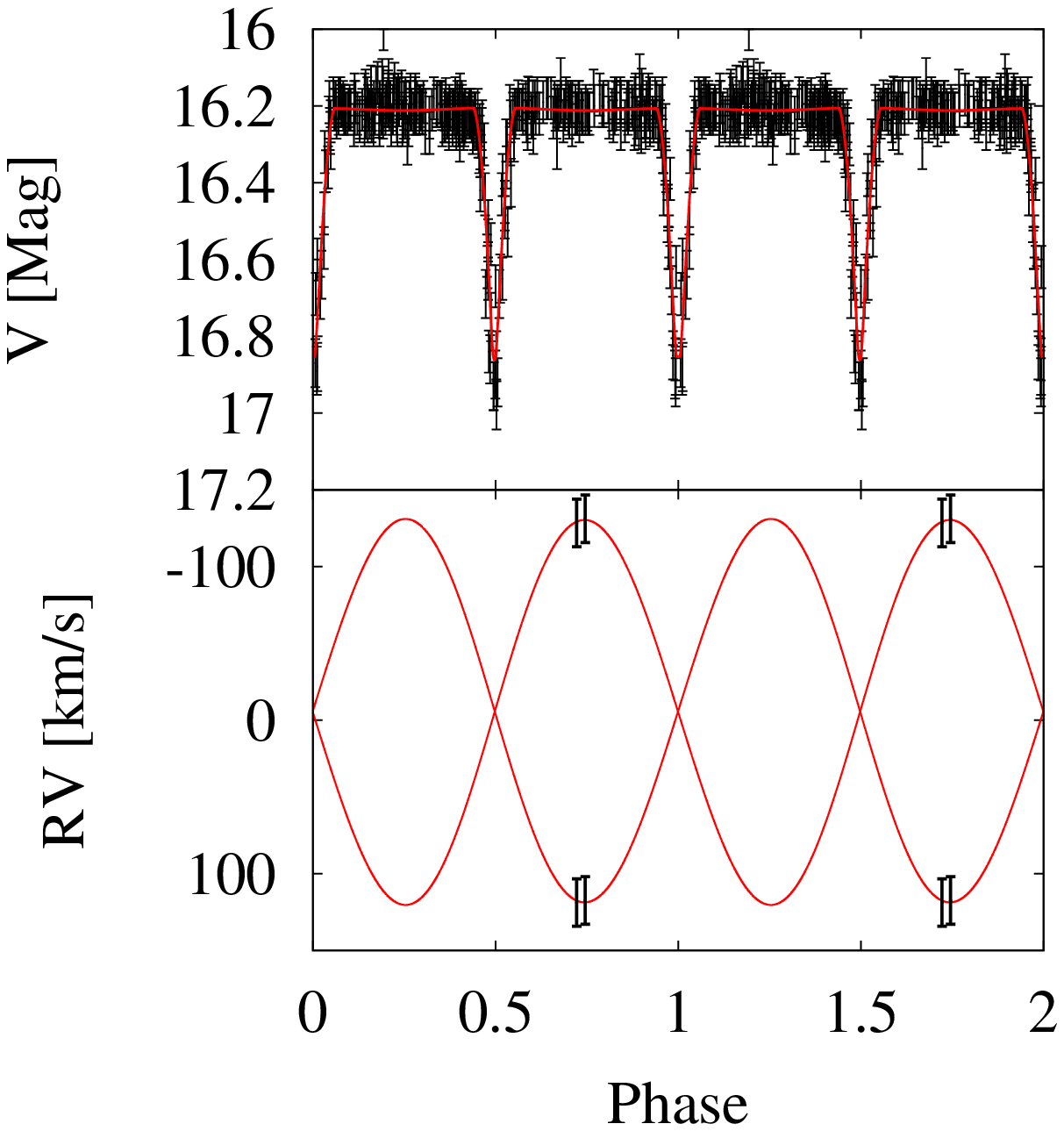}
  \caption{Joint light curve and RV analysis of CSSJ074118.8+311434 (left), CSSJ092128.3+332558 (middle), and CSSJ072108.8+344808 (right) using JKTEBOP.}
  \label{fig.fit}
\end{figure*}

With both the light curves and RV information in hand, we can determine the mass and the radius
of the system. We use JKTEBOP\footnote{http://www.astro.keele.ac.uk/jkt/codes/jktebop.html} \citep{2004MNRAS.351.1277S}, a
derivative of the the EBOP code originally developed by \cite{1981AJ.....86..102P}, with the capibility to
jointly fit the light and RV curves to determine the mass and radius of each stellar component, as well as
the parameters of the orbit. We fit the reference time of the primary eclipse (t$_0$), radius ratio of the
primary to the secondary (R$_2$/R$_1$), radius sum in terms of semi-major
axis ($\frac{R_1+R_2}{a}$), inclination angle ($i$), light ratio (L$_2$/L$_1$), orbital
eccentricity ($e$), augment of periastron ($\omega$), RV semi-amplitudes of the two
components (K$_1$ and K$_2$), and the systemaic velocity (V$_\mathrm{sys}$).
We use the best-fit results from DEBiL in Section \ref{sec.lc} as an initial guess for most of the parameters;
for semi-amplitudes K$_1$ and K$_2$, we assume the primary and secondary compoment are of similar mass, and assume the
RVs in Section \ref{sec.spec} are approximately to the maxumum values, to
estimate K$_1$ and K$_2$, respectively. The initial guess of the systemaic velocity is taken as the mean of the
estimated K$_1$ and K$_2$. The JKTEBOP fitting routine converges quickily (within 30 iterations),
indicating the DEBiL results provides a good initial guess. 
Our best-fit JKTEBOP results are showin in Fig. \ref{fig.fit} (in red colour) and
Table \ref{tab.fit}. We provide a detail discussion of each eclipsing binaries in the following section.

\begin{table}[!h]
\centering
\caption{Best-fit parameters for our eclipsing binary systems}
\begin{tabular}[t]{lrrr}
  \hline
  \hline
  Parameter & CSSJ0741 & CSSJ0921 & CSSJ0721 \\
  \hline
  & \multicolumn{3}{c}{{\it System parameters}} \\
  R.A. & 07:41:18.81 & 09:21:28.32 & 07:21:08.81 \\
  Dec. & +31:14:34.2 & +33:25:58.4 & +34:48:08.7 \\
  Period [day] & 1.30223 & 0.42647 & 0.404832 \\
  V [mag] & 15.8 & 14.06 & 16.27\\
  Spectral type & M3 & M2 & M3 \\
  RV [km/s] & 121.5 & 4.8 & -10.2 \\
  \hline
  & \multicolumn{3}{c}{{\it Modelled parameters}} \\
t$_0$ [MJD] & 5849.469(1) & 5322.1937(5) & 4450.3046(3)\\
(R$_1$+R$_2$)/$a$ & 0.24(5) & 0.53(7) & 0.31(1)\\
R$_2$/R$_1$ & 1.58(2.99) & 0.71(1.54) & 1.16(7)\\
$i$ [deg] & 83.5(5.1) & 73.5(9.7) & 89.9(96.2)  \\
e cos$\omega$ & -0.001(2) & 0.001(3) & -0.005(1)\\
e sin$\omega$ & -0.1(1) & 0.05(8) & -0.02(2)\\
K$_1$ [km/s] & 89.8(9.3) & 101.37$^\dagger$ & 124.89$^\dagger$\\
K$_2$ [km/s] & 76.8(9.3) & 141.63$^\dagger$ & 125.40$^\dagger$\\
V$_\mathrm{sys}$ [km/s] & 119.5(6.7) & 5.38$^\dagger$ & -5.58$^\dagger$\\ 
\hline
& \multicolumn{3}{c}{{\it Derived parameters}} \\
M$_1$ [M$_\odot$] & 0.289 & 0.418 & 0.329\\ 
M$_2$ [M$_\odot$] & 0.338 & 0.300 & 0.328\\
R$_1$ [R$_\odot$] & 0.393 & 0.663 & 0.334\\
R$_2$ [R$_\odot$] & 0.620 & 0.478 & 0.389\\
a [AU] & 0.020 & 0.010 & 0.009 \\
\hline
\hline
\multicolumn{4}{l}{The errors are shown in the bracket (to the last digit).}\\
  \multicolumn{4}{l}{$\dagger$ Unable to determine the errors.}\\ 
\end{tabular}
\label{tab.fit}
\end{table}

\section{Discussion}
\label{sec.dis}

\subsection{CSSJ074118.8+311434}
The best-fit model from JKTEBOP indicates M$_1$=0.289~M$_\odot$, M$_2$=0.338~M$_\odot$, R$_1$=0.393~R$_\odot$, 
and R$_2$=0.62~R$_\odot$. While the primary component is in good agreement with the emperical mass-radius 
relation of \cite{2009A&A...505..205D}, the secondary component's radius appears to be inflated. 
The inflated secondary stellar component might originate from tidal locking effect exerted by the
close companion, inducing enhanced stellar activity and inhibited convection, commonly seen in short period
binaries \citep{2011ApJ...728...48K}.

\subsection{CSSJ092128.3+332558}

The best-fit model from JKTEBOP indicates M$_1$=0.716~M$_\odot$, M$_2$=0.418~M$_\odot$, R$_1$=0.299~R$_\odot$, 
and R$_2$=0.663~R$_\odot$. Due to the limited observation time, we only obtained radial velocity measurements
at phase $\sim$0.25, hence JKTEBOP cannot provide reasonable error estimate of K$_1$, K$_2$, and V$_\mathrm{sys}$. Nevertheless we note that the best-fit system velocity
(5.38km/s) is in good agreement with the system velocity from LAMOST single-exposure (4.8km/s), indicating our best-fit results are in the right ballpark.
For CSSJ092128.3+332558, both stellar components show enlarged radius compared to the mass radius relation of \cite{2009A&A...505..205D}, indicating both compontent
are in tidal lock.

\subsection{CSSJ072108.8+344808}

The best-fit model from JKTEBOP indicates M$_1$=0.329~M$_\odot$, M$_2$=0.328~M$_\odot$, R$_1$=0.334~R$_\odot$, and R$_2$=0.389~R$_\odot$. Due to the limited observation time, we only obtained RVs at phase $\sim$0.75, so that were
not able to obtain error estimate from JKTEBOP. Nevertheless we note that the best-fit
system velocity (-5.58km/s) is consistent with the system velocity from LAMOST
single-exposure (-10.2km/s), indicating our best-fit results are in the right
ballpark. For CSSJ072108.8+344808, both stellar components are in good agreement
with the mass-radius relation of \cite{2009A&A...505..205D}.

\section{Summary and prospects}
\label{sec.sum}
We present a preliminary analysis of three double-lined M dwarf eclipsing binary systems
discovered by cross-matching eclipsing binaries charted
by Catalina Sky Survey and spectroscopic M dwarves in LAMOST. We obtained follow-up, medium resolution
spectra using GMOS on-board Gemini telescope, enabling us to disentangle the emission lines from both stellar
components, and providing us the RV information during the RV maxiumu states.
Our best-fit results suggest that both components of each eclipsing binaries are composed of M dwarves, and one of the stellar component of CSSJ074118.8+311434 even falls in the very low mass star regime (M$<$0.3 M$_\odot$).
warrant further follow-ups.
Due to the limited amount of available observation time, we were only able to obtain RV information
at one quadrature for CSSJ092128.3+332558 and CSSJ072108.8+344808.
Nevertheless their best-fit systematic velocities are in consistent with the RV from LAMOST single-shot
spectra, indicating our best-fit parameters are in the right ballpark.
Future high resolution spectroscopic observations will help pinning down the uncertainties of the fundamental
parameters of these systems. Multi-epoch spectra, focusing on H$\alpha$ or other indicator of stellar activities,
will also shed light on the strength of stellar activity, as a means to test the tidal-locking induced inflation scenario.

\begin{acknowledgements}
Guoshoujing Telescope (the Large Sky Area Multi-Object Fiber Spectroscopic Telescope LAMOST) is a National Major Scientific Project built by the Chinese Academy of Sciences. Funding for the project has been provided by the National Development and Reform Commission. LAMOST is operated and managed by the National Astronomical Observatories, Chinese Academy of Sciences.

The CSS survey is funded by the National Aeronautics and Space Administration under Grant No. NNG05GF22G issued through the Science Mission Directorate Near-Earth Objects Observations Program. The CRTS survey is supported by the U.S.-National Science Foundation under grants AST-0909182.

Based on observations obtained at the Gemini Observatory and processed using the Gemini IRAF package,
which is operated by the Association of Universities for Research in Astronomy, Inc., under a cooperative
agreement with the NSF on behalf of the Gemini partnership: the National Science Foundation (United States),
the National Research Council (Canada), CONICYT (Chile), Ministerio de Ciencia,
Tecnolog\'{i}a e Innovaci\'{o}n Productiva (Argentina), and Minist\'{e}rio da Ci\^{e}ncia, Tecnologia e Inova\c{c}\~{a}o (Brazil).

The authors wish to recognize and acknowledge the very significant cultural role and reverence that the summit of Maunakea
has always had within the indigenous Hawaiian community.  We are most fortunate to have the opportunity to conduct observations from this mountain.
\end{acknowledgements}

\label{lastpage}


\begin{thebibliography}{99}

\bibitem[\protect\citeauthoryear{Allard et al.}{2012}]{2012EAS....57....3A} Allard F., Homeier D., Freytag B., Sharp C.~M., 2012, EAS, 57, 3
 
\bibitem[Bai et al.(2016)]{2016RAA....16g...7B} Bai, Y., Luo, A.-L., Comte, G., et al.\ 2016, Research in Astronomy and Astrophysics, 16, 007 

\bibitem[\protect\citeauthoryear{Bertin \& Arnouts}{1996}]{1996A&AS..117..393B} Bertin E., Arnouts S., 1996, A\&AS, 117, 393 

\bibitem[\protect\citeauthoryear{Covey et al.}{2007}]{2007AJ....134.2398C} Covey K.~R., et al., 2007, AJ, 134, 2398 

\bibitem[Cui et al.(2012)]{2012RAA....12.1197C} Cui, X.-Q., Zhao, Y.-H., Chu, Y.-Q., et al.\ 2012, Research in Astronomy and Astrophysics, 12, 1197 

\bibitem[\protect\citeauthoryear{Demory et al.}{2009}]{2009A&A...505..205D} Demory B.-O., et al., 2009, A\&A, 505, 205 
  
\bibitem[Devor(2005)]{2005ApJ...628..411D} Devor, J.\ 2005, \apj, 628, 411 

\bibitem[\protect\citeauthoryear{Drake et al.}{2009}]{2009ApJ...696..870D} Drake A.~J., et al., 2009, ApJ, 696, 870 

\bibitem[\protect\citeauthoryear{Drake et al.}{2013}]{2013ApJ...763...32D} Drake A.~J., et al., 2013, ApJ, 763, 32 

\bibitem[\protect\citeauthoryear{Drake et al.}{2014}]{2014ApJS..213....9D} Drake A.~J., et al., 2014, ApJS, 213, 9 

\bibitem[\protect\citeauthoryear{Dressing \& Charbonneau}{2013}]{2013ApJ...767...95D} Dressing C.~D., Charbonneau D., 2013, ApJ, 767, 95 

\bibitem[Guo et al.(2015)]{2015RAA....15.1182G} Guo, Y.-X., Yi, Z.-P., Luo, A.-L., et al.\ 2015, Research in Astronomy and Astrophysics, 15, 1182 

\bibitem[\protect\citeauthoryear{Hook et al.}{2004}]{2004PASP..116..425H} Hook I.~M., J{\o}rgensen I., Allington-Smith J.~R., Davies R.~L., Metcalfe N., Murowinski R.~G., Crampton D., 2004, PASP, 116, 425 

\bibitem[\protect\citeauthoryear{Kraus et al.}{2011}]{2011ApJ...728...48K} Kraus A.~L., Tucker R.~A., Thompson M.~I., Craine E.~R., Hillenbrand L.~A., 2011, ApJ, 728, 48 

\bibitem[\protect\citeauthoryear{L{\'e}pine \& Gaidos}{2011}]{2011AJ....142..138L} L{\'e}pine S., Gaidos E., 2011, AJ, 142, 138 

\bibitem[\protect\citeauthoryear{Popper \& Etzel}{1981}]{1981AJ.....86..102P} Popper D.~M., Etzel P.~B., 1981, AJ, 86, 102 

\bibitem[Press et al.(1992)]{1992nrfa.book.....P} Press, W.~H., Teukolsky, 
S.~A., Vetterling, W.~T., 
\& Flannery, B.~P.\ 1992, Cambridge: University Press, |c1992, 2nd ed.,  

\bibitem[Seager \& Mall{\'e}n-Ornelas(2003)]{2003ApJ...585.1038S} Seager, S., \& Mall{\'e}n-Ornelas, G.\ 2003, \apj, 585, 1038 

\bibitem[\protect\citeauthoryear{Southworth, Maxted, \& Smalley}{2004}]{2004MNRAS.351.1277S} Southworth J., Maxted P.~F.~L., Smalley B., 2004, MNRAS, 351, 1277 

\bibitem[\protect\citeauthoryear{Thompson et al.}{2013}]{2013PASP..125..809T} Thompson M.~A., et al., 2013, PASP, 125, 809 

\bibitem[\protect\citeauthoryear{West et al.}{2011}]{2011AJ....141...97W} West A.~A., et al., 2011, AJ, 141, 97 

\bibitem[\protect\citeauthoryear{York et al.}{2000}]{2000AJ....120.1579Y} York D.~G., et al., 2000, AJ, 120, 1579 

\bibitem[\protect\citeauthoryear{Zhou et al.}{2015}]{2015MNRAS.451.2263Z} Zhou G., et al., 2015, MNRAS, 451, 2263 

\end{thebibliography}
\end{document}